\def\year{2021}\relax
\documentclass[letterpaper]{article} 
\usepackage{aaai21}  
\usepackage{times}  
\usepackage{helvet} 
\usepackage{courier}  
\usepackage[hyphens]{url}  
\usepackage{graphicx} 
\usepackage[switch]{lineno}
\usepackage{subfigure}
\usepackage{bm}
\usepackage{booktabs}
\usepackage{amsthm,amsmath,amssymb}
\usepackage{natbib}  
\usepackage{caption} 
\newcommand{\eg}{\emph{e.g.}}
\newcommand{\ie}{\emph{i.e.}}
\newcommand{\etc}{\emph{etc}}

\usepackage{algorithm}  
\usepackage{algorithmic}  

\title{SongMASS: Automatic Song Writing with Pre-training and Alignment Constraint}
\author{
     Zhonghao Sheng$^{1,}$\footnote{The first two authors contribute equally to this work.}, 
     Kaitao Song$^{2,*}$, 
     Xu Tan\footnote{Corresponding Author: Xu Tan, xuta@microsoft.com, and Wei Ye, wye@pku.edu.cn.}$^3$,
     Yi Ren$^4$,
     Wei Ye$^1$,
     Shikun Zhang$^1$, 
     Tao Qin$^3$
     \\
}
\affiliations{
    National Engineering Research Center for Software Engineering, Peking University$^1$, \\
    Nanjing University of Science and Technology$^2$,
    Microsoft Research Asia$^3$,
    Zhejiang University$^4$
    \\
    \{zhonghao.sheng,wye,zhangsk\}@pku.edu.cn, kt.song@njust.edu.cn, \{xuta, taoqin\}@microsoft.com, rayeren@zju.edu.cn
}

\begin{document}
\maketitle

\begin{abstract}
Automatic song writing aims to compose a song (lyric and/or melody) by machine, which is an interesting topic in both academia and industry. In automatic song writing, lyric-to-melody generation and melody-to-lyric generation are two important tasks, both of which usually suffer from the following challenges: 1) the paired lyric and melody data are limited, which affects the generation quality of the two tasks, considering a lot of paired training data are needed due to the weak correlation between lyric and melody; 2) Strict alignments are required between lyric and melody, which relies on specific alignment modeling. In this paper, we propose SongMASS to address the above challenges, which leverages masked sequence to sequence (MASS) pre-training and attention based alignment modeling for lyric-to-melody and melody-to-lyric generation. Specifically, 1) we extend the original sentence-level MASS pre-training to song level to better capture long contextual information in music, and use a separate encoder and decoder for each modality (lyric or melody);  2) we leverage sentence-level attention mask and token-level attention constraint during training to enhance the alignment between lyric and melody. During inference, we use a dynamic programming strategy to obtain the alignment between each word/syllable in lyric and note in melody. We pre-train SongMASS on unpaired lyric and melody datasets, and both objective and subjective evaluations demonstrate that SongMASS generates lyric and melody with significantly better quality than the baseline method without pre-training or alignment constraint. 
\end{abstract}

\section{Introduction}
\begin{figure}[!t]
\centering
\includegraphics[width=0.48\textwidth]{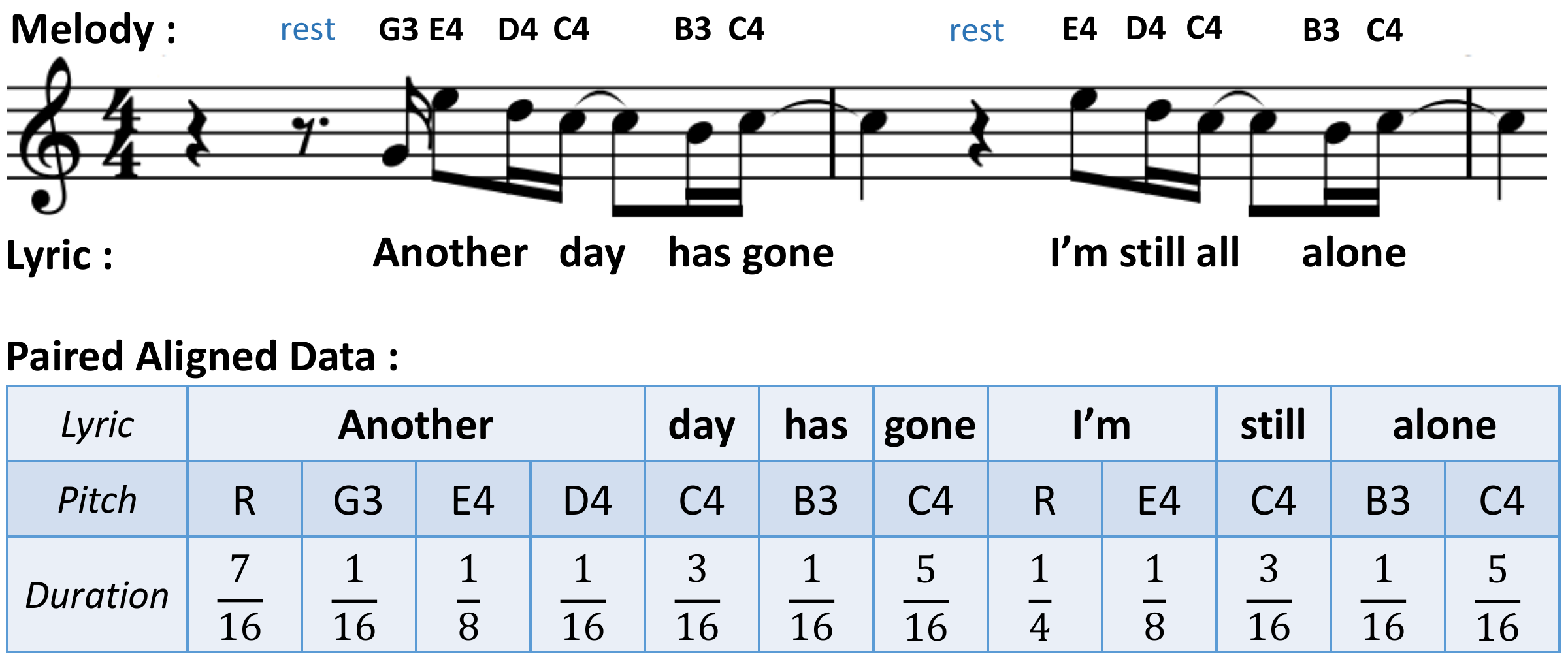}
\caption{A song fragment ``Another day has gone, I’m still all alone'' with its melody. The table shows the alignment of the lyric and melody (pitch and duration).}
\label{paired_data}
\end{figure}

Automatic song writing is an interesting and challenging task in both research and industry. Two most important tasks in automatic song writing are lyric-to-melody generation (L2M)~\cite{hangbo2019NMCLyrics,Yi2019ConditionalLSTMGAN,lee2019icomposer} and melody-to-lyric generation (M2L)~\cite{Watanabe2018melodylm,Xu2019Syllable,lee2019icomposer}. L2M and M2L can be regarded as sequence to sequence learning tasks and can be modeled by the techniques in natural language processing since both melody and lyric can be represented as discrete token sequence. However, L2M and M2L have distinctive characteristics that differ them from other sequence to sequence learning tasks: 1) lyric and melody are weakly correlated in L2M and M2L while in other tasks~\cite{bahdanau2014nmt,rush2015neural}, source and target sequences are strongly correlated in semantics; 2) one word or syllable in lyric always strictly aligns with one or more notes in melody, while other tasks do not require strict alignments. An example of aligned lyric and melody piece is shown in Figure~\ref{paired_data}.

The above distinctive characteristics throw several challenges in modeling L2M and M2L: 1) They require large amount of paired melody and lyric data to learn the mapping relationship between lyric and melody due to weak correlation. However, it is difficult to collect such large amount of paired data, and thus both tasks suffer from limited paired data; 2) They need additionally generate strict alignments between word/syllable in lyric and note in melody, and thus how to model the alignments well is critical to ensure the generation quality of lyric and melody. Previous works~\cite{hangbo2019NMCLyrics,Piji2020SongNet,Watanabe2018melodylm,lee2019icomposer} on L2M and M2L have not considered the scenario of limited paired data, and only leverage some greedy decisions for lyric and melody alignment, which cannot well address these challenges. In this paper, we propose SongMASS, an automatic song writing system for L2M and M2L, which addresses the first challenge with masked sequence to sequence pre-training and the second challenge with attention based alignment constraint.

Specifically, to handle the challenges of limited paired data, we leverage self-supervised pre-training on large amount of unpaired lyric and melody data. Since L2M and M2L are both sequence to sequence learning tasks, we adopt masked sequence to sequence pre-training (MASS)~\cite{kaitao2019MASS}, which is a popular pre-training method by masking a segment in the input sequence and predicting this segment in the output using an encoder-decoder framework. However, simply using original MASS in L2M and M2L cannot well handle the long lyric and melody sequence in song level and the diversity between lyric and melody modality. Therefore, we introduce two extensions on MASS: 1) Instead of masking in a single sentence in original MASS, we design a song-level masked pre-training strategy to capture longer contextual information, since music usually has repeat structure and relies on long context. 2) Unlike original MASS, we use separate encoder-decoder for lyric-to-lyric and melody-to-melody masked pre-training since they are in different modalities. However, separate training of lyric-to-lyric and melody-to-melody cannot ensure to learn a shared latent space between lyric and melody and thus could harm the transformation between them. Therefore, we add supervised training with paired lyric and melody data to guide the pre-training towards learning a shared latent representation between lyric and melody modality. 

To address the challenges of lyric-melody alignment, we propose to align the word/syllable in lyric and note in melody based on the encoder-decoder attention. Due to long melody and lyric sequence in a song, we split the alignment into sentence level and token level. To ensure sentence-level alignment, we constrain each sentence in target sequence to only attend to the corresponding sentence in source sequence during training and inference. We add an additional [SEP] token as the sentence boundary in each sequence, and during inference, once a [SEP] token is predicted in the target side, we switch the attention to the next source sentence. For token-level alignment, we add constraints on the attention matrix using the ground-truth alignment in the paired training data during training, and use a dynamic programming algorithm on the generated attention matrix during inference to obtain the final alignments. 

The contributions of our method are as follows:
\begin{itemize}
    \item We are the first to leverage pre-training to address the low-resource challenge on L2M and M2L, by introducing two extensions on MASS including song-level masked pre-training and using supervised pre-training to guide the separate encoder-decoder of lyric and melody to the same latent space. 
    \item To handle the alignment between word/syllable in lyric and note in melody, we design the attention-based sentence-level and token-level alignment constraints and a dynamic programming algorithm to obtain precise alignments. 
    \item Experimental results with objective and subjective evaluations demonstrate that SongMASS significantly improves the quality of lyric and melody generation with the help of pre-training and alignment constraint.
\end{itemize}

\section{Background}

\paragraph{Automatic Song Writing}
Automatic song writing usually covers several tasks including lyric generation~\cite{Eric2015Dope}, melody generation~\cite{Zhu2018XiaoIce}, lyric-to-melody generation (L2M)~\cite{Keunwoo2016TextMC,Yi2019ConditionalLSTMGAN} and melody-to-lyric generation (M2L)~\cite{hangbo2019NMCLyrics,Piji2020SongNet}. In this work, we focus on L2M and M2L. \citet{Keunwoo2016TextMC,Yi2019ConditionalLSTMGAN} generated melody conditioned on the lyrics with RNN-based language model. \citet{lee2019icomposer,hangbo2019NMCLyrics} used sequence to sequence model for L2M and M2L. However, these works on L2M and M2L usually only used limited paired data, without leveraging large-scale unpaired data. On the other hand, some works only focused on L2M and M2L on the sentence level, assuming there are strict one-to-one mapping in the training data, which cannot compose a complete song. Some other works~\cite{hangbo2019NMCLyrics,Watanabe2018melodylm} explicitly predicted the alignment flag (e.g., whether switches to next word/syllable when predicting notes or not) in the model, with a greedy decision in the word/syllable or note level, which is not flexible and fail to capture the global alignment in the whole sentence. In this paper, we propose SongMASS, which uses sequence to sequence pre-training method to leverage the unpaired lyric and melody data, and attention-based alignment constraints for global and precise lyric-melody alignment. 

\paragraph{Pre-training Methods} 
\label{background_ptm}
Pre-trained language models (\eg, BERT~\cite{Devlin2019Bert}, GPT~\cite{radford2018gpt}, XLNet~\cite{Yang2019XLNet}, MASS~\cite{kaitao2019MASS} and \etc) have achieved significant progress in natural language processing. They usually employ specific self-supervised tasks and pre-train on large-scale unlabeled data corpus to improve the understanding and generation capability. MASS~\cite{kaitao2019MASS} is the first and one of most successful pre-training methods for sequence to sequence learning tasks, and several pre-training methods such as BART~\citep{lewis2019bart} and T5~\citep{raffel2019exploring} are also proposed to handle this kind of task. In this paper, we build our pre-training method upon MASS considering its popularity for sequence to sequence learning tasks and suitability for different modalities. Given a sequence from the unpaired sentence corpus, MASS randomly replaces a segment of tokens with mask tokens and takes the masked sequence as the encoder input and predicts the masked segment in the decoder. We leverage the basic idea of MASS and extend it with several improvements to address the distinctive challenges in the pre-training of L2M and M2L.

\paragraph{Alignment Modeling}
Alignment modeling builds the correlation between the tokens in source and target sequences, which plays an important role in sequence to sequence tasks. In L2M and M2L tasks, previous works usually used greedy alignment mechanisms to handle the correlation between lyric and melody. For example, \citet{Watanabe2018melodylm} used the Needleman-Wunsch algorithm~\cite{Needleman-Wunsch} to count the alignment of lyric and melody. \citet{hangbo2019NMCLyrics} predicted how many syllables in the predicting word given current note input. However, these greedy alignment strategies cannot provide flexible and global alignments in the sentence level. In other sequence to sequence learning tasks like neural machine translation, \citet{bahdanau2014nmt,luong2015effective} introduced attention mechanism to learn the relationship between source and target languages. In this paper, we leverage the attention mechanism to build the global and soft alignment between lyric and melody, and finally design a dynamic programming method to obtain the strict alignment between word/syllable and note.

\section{Method}

\subsection{System Overview}
\begin{figure}[h]
    \centering
    \includegraphics[width=0.45\textwidth]{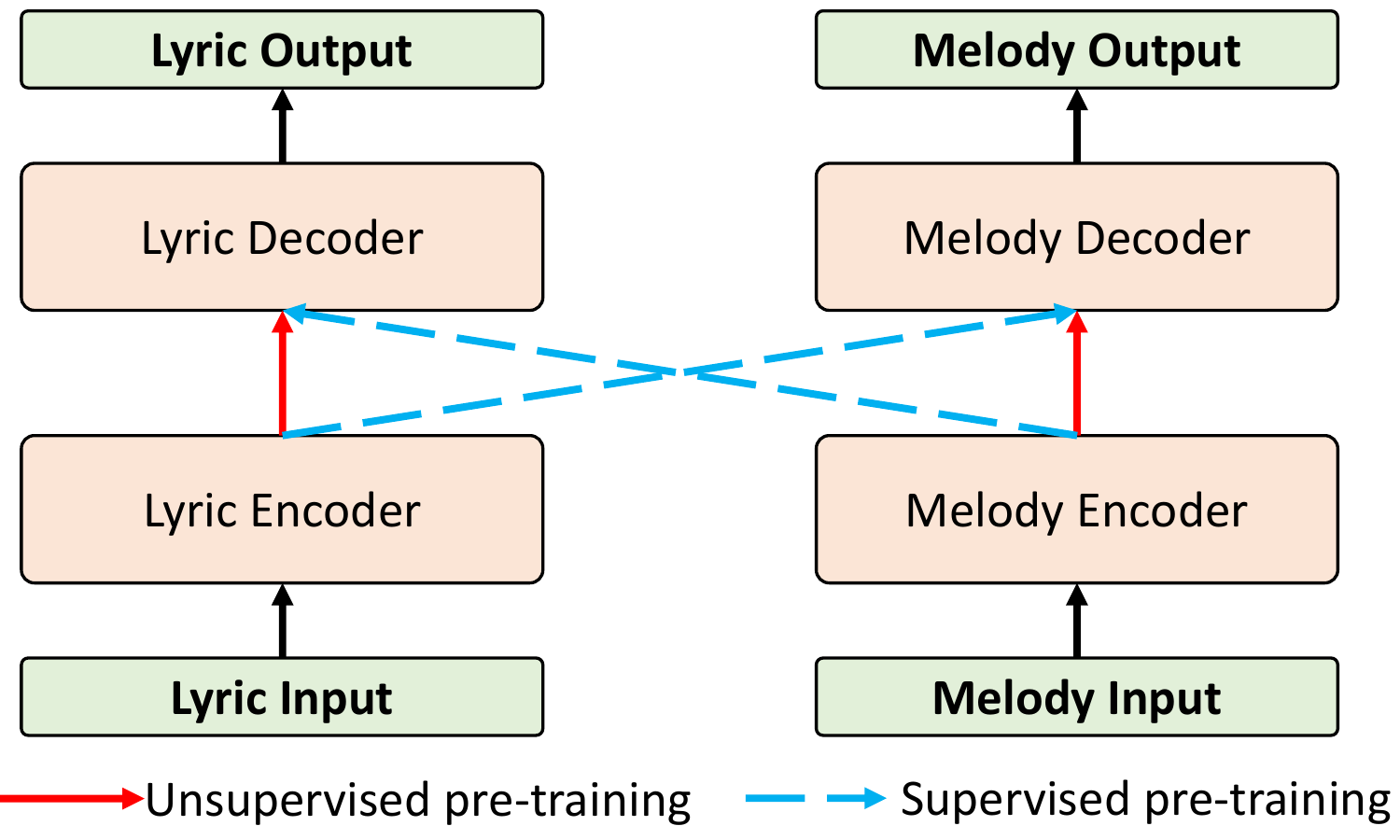}
    \caption{\label{arch} The overall architecture of our SongMASS framework. The red line means unsupervised pre-training on lyric-to-lyric or melody-to-melody. The blue dotted line is supervised pre-training on lyric-to-melody or melody-to-lyric.}
\end{figure}

The overall architecture of SongMASS for L2M and M2L is shown in Figure~\ref{arch}, which adopts the Transformer~\cite{Vaswani2017Attention} based encoder-decoder framework. We employ separated encoders and decoders for lyric and melody respectively due to the large diversity between lyric and melody. To leverage the knowledge from large-scale unlabeled lyrics or melodies, we perform MASS pre-training for lyric-to-lyric and melody-to-melody in our framework. We pre-train our model in song level to better capture long contextual information for lyric or melody sequence, and incorporate supervised learning (lyric-to-melody and melody-to-lyric) into our pre-training to learn a shared latent space between different modalities. To learn the alignment between word/syllable in lyric and note in melody, we further leverage sentence-level constraint and token-level constraint into our model to guide the alignment between lyric and melody. We use a dynamic programming strategy to obtain the final strict alignment between the lyric and melody. In below, we describe the details of our pre-training methods and alignment strategies in SongMASS.

\begin{figure*}[h]
\centering
\includegraphics[width=0.98\textwidth]{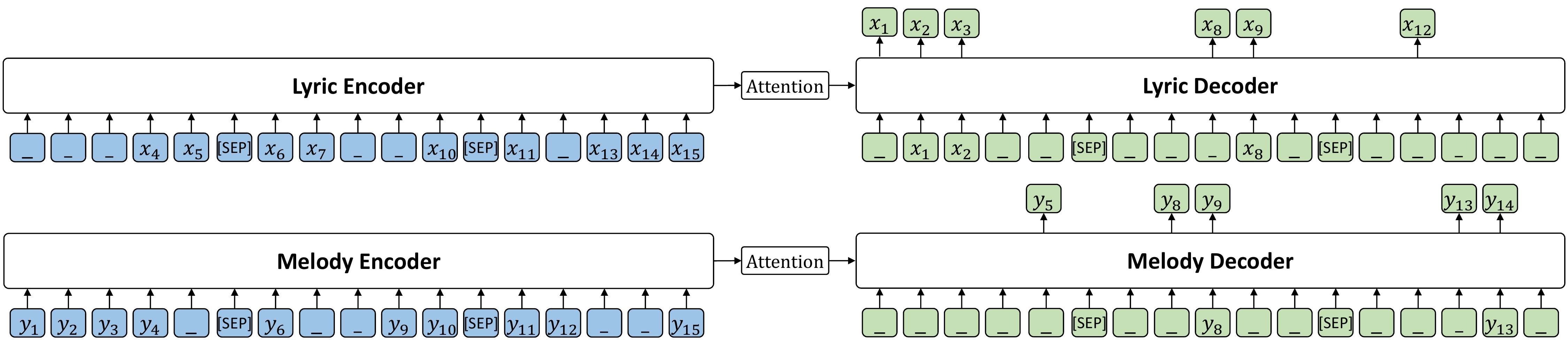}
\caption{\label{song_level_mass} The song-level MASS pre-training.}
\end{figure*}

\subsection{Pre-training Methods}
In this subsection, we introduce our pre-training methods, including song-level MASS pre-training to capture long contextual information from the whole song, and supervised pre-training to learn a shared latent space between lyric and melody modality.

\paragraph{Song-Level MASS Pre-training} 
As mentioned in Section~\ref{background_ptm}, the original MASS pre-training is designed to help the model understand and generate sequence in sentence level. However, instead of using sentence-level information, we expect model to capture long contextual information in song level (\ie, the whole song). Therefore, we introduce the song-level MASS pre-training to address this issue. 

Denote $\mathcal{X'}$ and $\mathcal{Y'}$ as the corpus of unpaired lyrics and melodies in the song level respectively. For any $x \in \mathcal{X'}$ and $y \in \mathcal{Y'}$, we split the song-level sequence into multiple sentences and insert a special token $\rm{[SEP]}$ in the boundary of adjacent sentences. For every sentence from the song-level sequence, we perform the same mask strategy as in the original MASS (as mentioned in Section~\ref{background_ptm}). The details of the masking strategy are shown in Figure~\ref{song_level_mass}. The encoder takes the masked song-level sequence as input and the decoder predicts masked fragments corresponding to all the sentences in this song. The formulation of song-level MASS is as follows:
\begin{equation}
\small
    \label{eq_unsup}
    \begin{aligned}
        & L(\mathcal{X};\theta^{enc},\theta^{dec})   = \sum_{x\in\mathcal{X}} \sum_{i=1}^S \log P(x^{u_i:v_i}|x^{\backslash \{ u_i:v_i\}^S_{i=1}};\theta^{enc},\theta^{dec})\\
        & = 
        \sum_{x\in\mathcal{X}} \sum_{i=1}^S \log \prod_{t=u_i}^{v_i}  P(x^{u_i:v_i}_{t}| x^{u_i:v_i}_{<t}, x^{\backslash \{ u_i:v_i\}^S_{i=1}};\theta^{enc},\theta^{dec}),\\
    \end{aligned}
\end{equation}
where $S$ represents the number of sentences in sequence $x$, $x^{\backslash \{ u_i:v_i\}^S_{i=1}}$ represents the masked song-level sequence, and $x^{u_i:v_i}$ represents the masked segment in the $i$-th sentence. We define $\theta_{x}^{enc}$, $\theta_{x}^{dec}$, $\theta_{y}^{enc}$, $\theta_{y}^{dec}$ as the parameters of lyric encoder, lyric decoder, melody encoder and melody decoder. The loss for lyric-to-lyric generation is $L_{x}=L(\mathcal{X};\theta^{enc}_{x}, \theta^{dec}_{x})$ and the loss for melody-to-melody is $L_{y}=L(\mathcal{Y};\theta^{enc}_{y}, \theta^{dec}_{y})$.

\paragraph{Supervised Pre-training} Although MASS pre-training can help the model understand and generate lyric and melody respectively, the model cannot learn to generate melody from lyric and lyric from melody. What is worse is that the encoder-decoder models for lyric and melody cannot align in the same latent space and may deviate from each other, which will harm the transformation between lyric and melody. To prevent them from deviating and help align them together, we leverage the supervised training on lyric-melody paired data in the pre-training process. Given paired corpus $(\mathcal{X}, \mathcal{Y})$, the loss of the supervised pre-training is
\begin{equation}
\small
    \begin{aligned}
        {L}(\mathcal{X},\mathcal{Y};  \theta^{enc},&\theta^{dec})       = \sum_{(x,y)\in(\mathcal{X},\mathcal{Y})}\log P(y|x; \theta^{enc},\theta^{dec})\\
        & = \sum_{(x,y)\in(\mathcal{X},\mathcal{Y})}\log \prod_{t=1}^{|y|}  P(y_{t}|y_{<t},x; \theta^{enc},\theta^{dec}).\\
    \end{aligned}
\end{equation}
The supervised pre-training is applied on both lyric-to-melody and melody-to-lyric generation. The loss for lyric-to-melody generation is $L_{xy}=L(\mathcal{X}, \mathcal{Y};\theta^{enc}_{x}, \theta^{dec}_{y})$ and the loss for melody-to-lyric is $L_{yx}=L(\mathcal{Y}, \mathcal{X};\theta^{enc}_{y}, \theta^{dec}_{x})$.

Finally, the total pre-training loss is
\begin{equation}
\small
\label{eq2}
L_{pt}=L_{x }+L_{y}+L_{xy}+L_{yx},
\end{equation}
where $L_x$ and $L_y$ are the unsupervised MASS pre-training loss described in Equation~\ref{eq_unsup}, and $L_{xy}$ and $L_{yx}$ are the supervised pre-training loss. During fine-tuning, we only use $L_{xy}$ for lyric-to-melody generation and $L_{yx}$ for melody-to-lyric generation.

\subsection{Alignment Strategy}
In this subsection, we describe how to learn the alignment between lyric and melody in SongMASS. The basic idea is to leverage the encoder-decoder attention to infer the alignment between each word/syllable in lyric and note in melody. In order to extract strict alignment, we explicitly add constraints on the attention to learn effective attention patterns during training and inference. Due to the long song-level sequence, we divide our alignment strategy into sentence-level constraint and token-level constraint.

\begin{figure}[!h]
\centering
\includegraphics[width=0.47\textwidth]{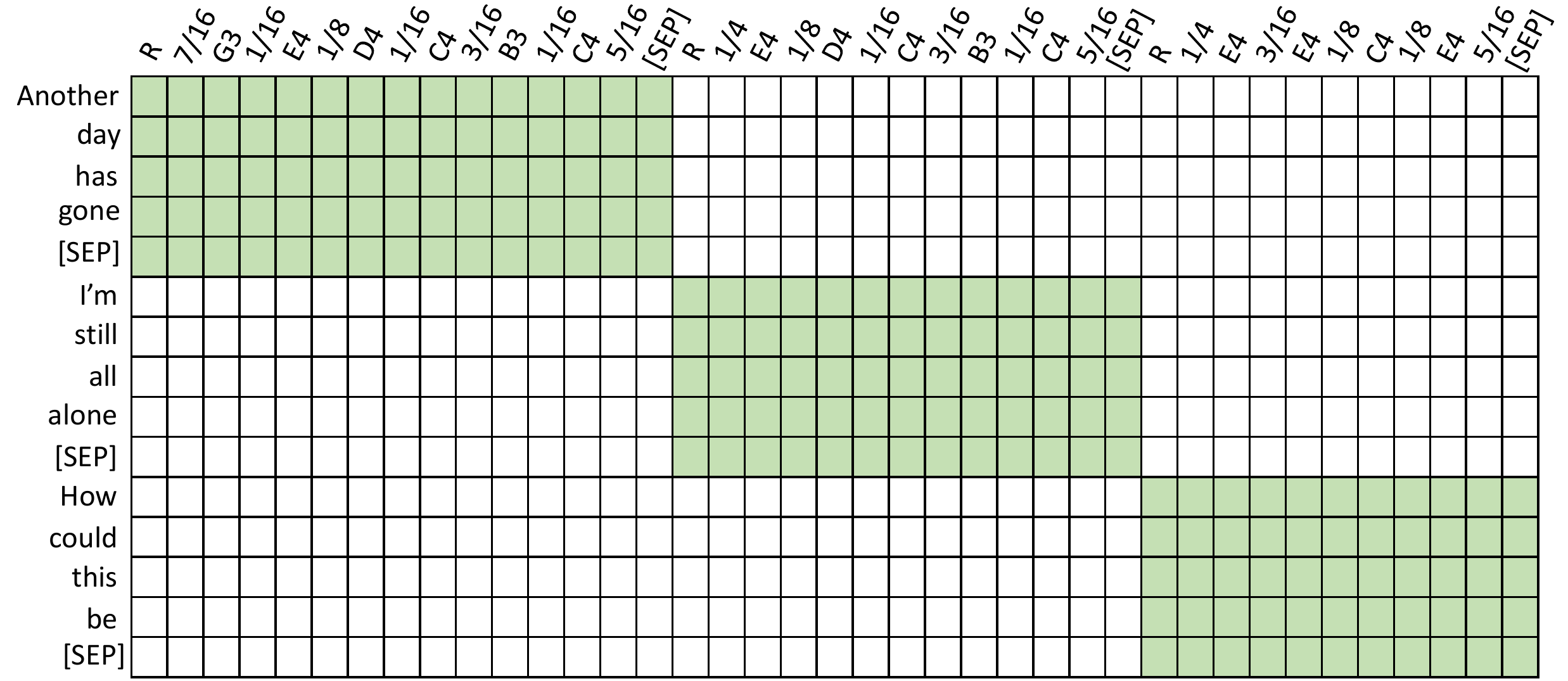}
\caption{\label{sent_mask} Sentence-level attention mask.}
\end{figure}

\paragraph{Sentence-Level Constraint} A song consists of multiple lyric sentences and melody phrases. In song writing, the given lyrics and melodies are naturally grouped into sentences or phrases. The lyric sentences and the melody phrases are strictly aligned in the training data. So we constrain that each sentence (lyric sentence or melody phrase) in the target sequence can only attend to the corresponding sentence (melody phrase or lyric sentence) in source sequence. Specifically, we apply a sentence-level constraint mask on the encoder-decoder attention. We denote $y_i$ and $x_j$ as the $i$-th token in the target sequence and $j$-th token in the source sequence respectively. We assume the representations of $x_j$ and $y_i$ from the previous Transformer layer as $h_j^{enc}$ and $h_i^{dec}$.
So the attention score between target token $y_i$ and source token $x_j$ is computed as:

\begin{equation}
\small
f(i, j) = \frac{h_i^{dec}W^{Q} (h_j^{enc}W^{K})^T}{\sqrt{d_z}} + M(i, j), \\
\end{equation}
\begin{equation}
\small
    A(i, j) = \frac{\exp f(i, j)} {\sum_j \exp f(i, j)}
\end{equation}
where $A(i, j)$ calculates the attention score between the $y_i$ and $x_j$. $W^Q$,$W^K$ $\in \mathbb{R}^{d_z \times d_z}$ are model parameters, and $d_z$ is the dimension of the hidden representations. $M(i, j)$ represents the mask element between $y_i$ and $x_j$, whose value is set as follows:
\begin{equation}
\small
M(i,j)=
    \begin{cases}
    0 & \rm{ID}(y_i) = \rm{ID}(x_j) \\
    -\infty & \rm{ID}(y_i) \ne \rm{ID}(x_j) 
    \end{cases}.
\end{equation}
where $\rm{ID}(x)$ gets the index of the sentence that the token $x$ belongs to. $M$ is used as our sentence-level alignment constraint, as shown in Figure \ref{sent_mask}. Besides, we insert a special token [SEP] in the sentence boundary of the input and output sequences as shown in Figure~\ref{song_level_mass}, to help the model better capture the sentence boundary information and identify which sentence of the input sequence should to be attended to at the current step during the inference stage. Benefiting from such design, we guarantee the number of sentences in the generated sequences is consistent with the input sequence.

\begin{figure}[!t]
\centering
\includegraphics[width=0.47\textwidth]{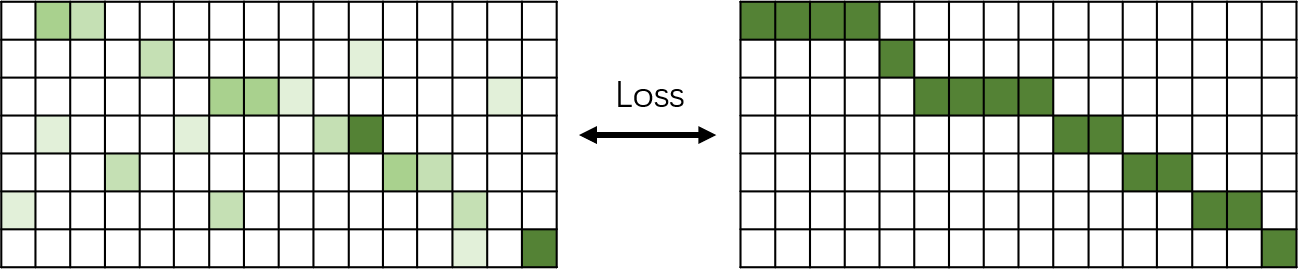}
\caption{\label{token_level_attn} Token-level guided attention mask.}
\end{figure}

\paragraph{Token-Level Constraint} Unlike sentence-level alignment, the alignment choices between each word/syllable and note are more flexible. Therefore, we propose a regularization term on the encoder-decoder attention during the training on paired data, and apply a dynamic programming algorithm on the attention matrix to obtain the final strict alignment during inference. We expect the attention weight between $y_i$ and $x_j$ to follow:
\begin{equation}
\small
u(i, j) = 
\begin{cases}
\frac{1}{T} & \text{if $y_i$ is aligned to $x_j$}, \\
0 & \text{Otherwise}, 
\end{cases}
\end{equation}
where $T$ is the number of tokens in the source sentence that $y_i$ is aligned to. As shown in Figure~\ref{token_level_attn}, we add a regularization term to constrain the attention weights: 
\begin{equation}
\small
    L_{att} = \frac{1}{N * M}\sum_{i=1}^{M}\sum_{j=1}^{N}\Vert A(i,j)-u(i, j)\Vert_2,
\end{equation}
where $\Vert\cdot\Vert$ represents L2-Norm. $N$ and $M$ are the number of tokens in the source and target sentence respectively. Finally, the loss function is:
\begin{equation}
\small
    L = L_{pt} + \alpha \cdot L_{att},
\end{equation}
where $\alpha$ is the hyper-parameter of $L_{att}$, and $L_{pt}$ is the pre-training loss defined in Equation~\ref{eq2}.

When all tokens in a sentence are generated and the attention matrix $A$ is obtained, we infer the global alignment by applying a dynamic programming algorithm as shown in Algorithm~\ref{dp}. We consider the following cases: a target token is aligned to one or many source tokens, and a source token is aligned to one or many target tokens. For the first case, as shown in Line 7 - 12 in Algorithm~\ref{dp}, we search for a $k$ that the alignment between $y_{i}$ and $x_{[K+1:j]}$ reaches the highest score, which is calculated by summing all the corresponding attention weights. Similarly for the second case, as shown in Line 13 - 18 in Algorithm~\ref{dp}, we search for a $k$ that the alignment between $y_{[k+1:i]}$ and $x_{[j]}$ reaches the highest score. We take the average weights as score in the second case, since the weights of the target sequence dimension are not normalized like that of the source sequence. We choose the higher score of the two cases and save the aligned pair. 

\begin{algorithm}[!t]
\small
	\caption{\label{dp} DP for Melody-Lyric Alignment}
	\begin{algorithmic}[1]
	    \STATE \textbf{Input}: Attention matrix $A \in \mathbb{R}^{N \times M}$, score matrix $F \in \mathbb{R}^{(N+1) \times (M+1)}$, path matrix $\text{Path}$, source sequence $x$ and target sequence $y$. $N$ and $M$ are the length of $x$ and $y$.
	    \STATE \textbf{Output}: The aligned pairs list $D$.
	    \STATE \textbf{Initialize}: F is initialized as $-\infty$. $\text{Path}$ is initialized as an empty matrix with a shape of $(N+1) \times (M+1)$.
	    \STATE {$F[0][0] = 0$}
	    \FOR{$i=1$ to $T$}
	    \FOR{$j=1$ to $S$}
	    \FOR{$k=0$ to $j-1$}
	    \STATE $\text{score} = F[i-1][k] + \sum_{h=k+1}^{j}A[i][h]$
	    \IF{$\text{score} \ge F[i][j]$}
	    \STATE $F[i][j] = \text{score}$, $\text{Path}[i][j]=(i-1,k)$
	    \ENDIF
	    \ENDFOR
	    \FOR{$k=0$ to $i-1$}
	    \STATE $\text{score} = F[k][j-1] + \sum_{h=k+1}^{i} \frac{A[h][j]}{i-k}$
	    \IF{$\text{score} \ge F[i][j]$}
	    \STATE $F[i][j] = \text{score}$, $\text{Path}[i][j] = (k,j-1)$.
	    \ENDIF
	    \ENDFOR
	    \ENDFOR
	    \ENDFOR
	    \STATE $m,n = M,N$
	    \WHILE {$m \ne 0$ \AND $n \ne 0$}
	    \STATE $i, j = \text{Path}[m][n]$
	    \STATE \text{add the aligned pair ($x_{[j+1:n]}$,$y_{[i+1:m]}$) to } D
	    \STATE $m, n = i, j$
	    \ENDWHILE
	    \RETURN $D$
	\end{algorithmic}
\end{algorithm}

\section{Experiments and Results}

\begin{table*}[h]
\centering
\begin{tabular}{l|cccc|c} 
    \toprule
    & \multicolumn{4}{c|}{Lyric-to-Melody} & Melody-to-Lyric  \\
    & PD ($\%$) $\uparrow$ & DD ($\%$) $\uparrow$ & MD $\downarrow$ & PPL $\downarrow$ & PPL $\downarrow$ \\
    \midrule
    Baseline & 38.20 & 52.00 & 2.92 & 3.27  & 37.50 \\
    \midrule 
    \textbf{SongMASS} & 57.00 & 65.90 & 2.28 & 2.41 & 14.66\\
     $-$ pre-training & 43.50 & 57.00 & 2.79 & 3.72 & 45.10 \\
     ~~~~$-$ separate encoder-decoder & 55.00 & 64.80  & 2.32 & 2.53 & 15.57 \\
     ~~~~$-$ supervised loss & 47.20 & 53.60  & 3.29 & 2.92 & 27.50 \\
     $-$ alignment & 56.10 & 65.20  & 2.36 & 2.07 & 8.54 \\
 
    \bottomrule
\end{tabular}
\caption{\label{main_result}Results of lyric-to-melody and melody-to-lyric generation in objective evaluation.}
\end{table*}

\subsection{Experimental Setup}
\paragraph{Dataset} 
\textit{Unpaired Lyric and Melody.} We use ``380,000+ lyrics from MetroLyrics"\footnote{\url{https://www.kaggle.com/gyani95/380000-lyrics-from-metrolyrics}} as our unpaired lyrics for pre-training, which contains 362,237 songs. The lyrics in each song are split into sentences by the line break. For unpaired melodies, we choose ``The Lakh MIDI Dataset"~\cite{raffel2016}\footnote{\url{https://colinraffel.com/projects/lmd}}. The dataset contains 176,581 MIDI files with complete tracks, and we extract the melody tracks by Midi-miner\footnote{https://github.com/ruiguo-bio/midi-miner}. Finally, we get 65,954 melodies as our unpaired data for pre-training.
According to the characteristics of vocal melody, we consider the pitch and duration tokens of each note as the melody sequence. Each melody is transposed to the scale of C major or A minor. All the notes are shifted by octave so that the most pitches of the song fall into one-lined octave (MIDI pitch from 60 to 71). For unpaired melody MIDI file, we calculate the starting beat and duration of the note based on the absolute time and the BPM (Beats Per Minute), all the notes are aligned to 1/16 notes as paired data. We spread the melodies into sequences of pitch-duration patterns, as melody sequences for our model. For example, the melody in Figure \ref{paired_data} will be represented as ``R, 7/16, G3, 1/16, E4, 1/8 ...". During pre-training, we simply split the unpaired melodies into phrases according to the average phrase length in paired data, since there is no natural phrase segmentation symbol in the MIDI files.
\textit{Paired Lyric and Melody.} We use the LMD dataset~\cite{Yi2019ConditionalLSTMGAN}\footnote{\url{https://github.com/yy1lab/Lyrics-Conditioned-Neural-Melody-Generation}} which contains aligned melodies and lyrics from 7,998 songs. We apply the same operation, as aforementioned, to process melody and lyric data. The lyrics/melodies are split into sentences/phrases based on the annotations.

\paragraph{Model Configuration and Training}
We choose Transformer~\citep{Vaswani2017Attention} as our basic model structure, which consists of 6 encoder/decoder layers. The hidden size and filter size of each layer are set as 512 and 2048. The number of attention heads is 8. We use the same masking strategy as in ~\citet{kaitao2019MASS}. We use Adam optimizer~\citep{Diederik2015Adam} with a learning rate of 5e-4. The model is trained on a NVIDIA Tesla T4 GPU card, and each mini-batch contains 4096 tokens. During training, we apply dropout with the rate of 0.1. The hyper-parameter $\alpha$ is set as 0.5. The dataset is split as training/valid/test set with a ratio of 8:1:1. Our baseline is a standard Transformer model, using the same model configuration with SongMASS but without any pre-training or alignment constraints.

\subsection{Evaluation Metrics}
\label{sec_metric}
In this subsection, we introduce the objective and subjective metrics used in this paper to evaluate the quality of lyric-to-melody and melody-to-lyric generation.
\paragraph{Objective Evaluation} We mainly measure the similarity between the generated melody and ground-truth melody in lyric-to-melody generation, in terms of pitch and duration distribution and melody sequence, which are described below. We use perplexity (PPL) to measure the model fitness for both lyric-to-melody and melody-to-lyric generations. Besides, we also use alignment accuracy to measure alignment quality in two generation tasks, which is also described below.
\begin{itemize}
\item \textit{PD} and \textit{DD} (Pitch and Duration Distribution Similarity): We calculate the distribution (frequency histogram) of pitches and durations in melodies, and measure the similarity (average overlapped area~\citep{DBLP:journals/corr/abs-2008-07703}) of the distribution between generated melodies and ground-truth melodies: $\frac{1}{N_{\text{s}}}\sum_{i=1}^{N_{\text{s}}}OA(\text{Dis}_i,\hat{\text{Dis}}_i)$, where $\text{Dis}_i$ and $\hat{\text{Dis}}_i$ represent the pitch or duration distribution of the $i$-th generated and ground-truth song respectively, $N_{s}$ is the number of songs in the testset, OA represents the average overlapped area.
\item \textit{MD} (Melody Distance): To evaluate the pitch trend of the melody, we spread out the notes into a time series of pitch according to the duration, with a granularity of 1/16 note. We subtract each pitch with the average pitch of the entire sequence for normalization. To measure the similarity between the generated and ground-truth time series with different lengths, we use dynamic time warping~\citep{berndt1994using} to measure their distance.

\item \textit{Alignment Accuracy}: To evaluate the alignment between melodies and lyrics, for each token in the source sequence, we calculate how many tokens in the target sequence (generated or ground-truth) are aligned to it, and check if the number of the tokens in the generated sequence equals to that in the ground-truth sequence. We calculate the ratio of equals among all source tokens and all songs in the test set to obtain the alignment accuracy.
\end{itemize}

\begin{figure*}[!h]
    \centering
    \subfigure[\label{Song_level_vis} Left: without sentence-level constraints. Right: with sentence-level constraints.]{
        \includegraphics[width=0.48\textwidth]{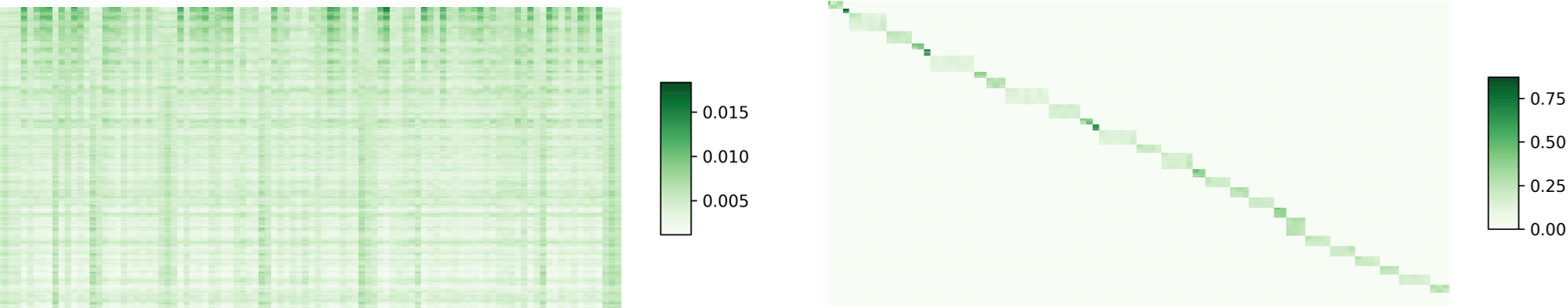}
    }
    \hspace{0.1cm}
    \subfigure[\label{Sentence_level_vis} Left: without token-level constraints. Right: with token-level constraints.]{
        \includegraphics[width=0.48\textwidth]{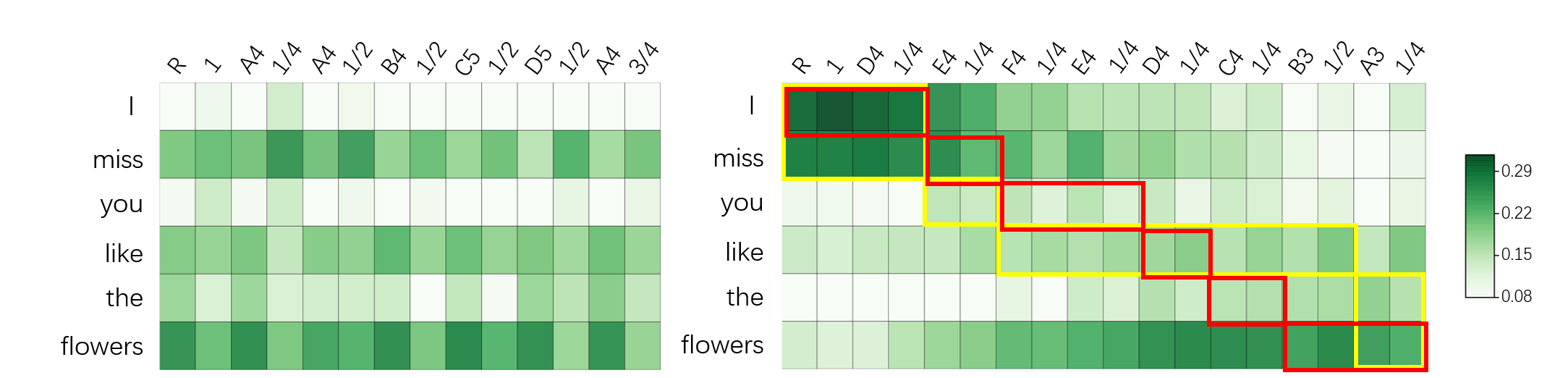}
    }
    \caption{\label{song_visualization} Attention visualization. All of the results are displayed on the average attention score of all heads in the last layer of the encoder-decoder attention in Transformer. In Figure~\ref{Sentence_level_vis}, the red blocks are the alignments searched by our dynamic programming algorithm while the yellow blocks are by the greedy algorithm described in the second paragraph in Section~\ref{sec_method_analysis}.}
\end{figure*}

\paragraph{Subjective Evaluation} 
For subjective evaluation, we invite 5 participants with professional knowledge in music and singing as human annotators to evaluate 10 songs (338 pairs of generated lyric sentences and melody phrases) randomly selected from our test set. We require each annotator to answer some questions using a five-point scale, from 1 (Poor) to 5 (Perfect). The whole evaluation is conducted in a blind-review mode. Inspired by \citet{Watanabe2018melodylm}, the metrics to evaluate the generated lyrics are as follows: 1) \textit{Listenability}: Is the lyric sounds natural with the melody? 2) \textit{Grammaticality}: It the lyric grammatically correct? 3) \textit{Meaning}: Is the lyric meaningful? 4) \textit{Quality}: What is the overall quality of the lyric? The metrics to evaluate the melody are as follows: 1) \textit{Emotion}~\cite{hangbo2019NMCLyrics}: Does the melody represent the emotion of the lyrics? 2) \textit{Rhythm}~\cite{Zhu2018XiaoIce}: Are the note durations and pauses of the melody sound natural? 3) \textit{Quality}~\cite{Watanabe2018melodylm}: What is the overall quality of the melody?

\begin{table}[h]
\centering
\begin{tabular}{lcc}
\toprule
  \textbf{Metric} & \textbf{Baseline} & \textbf{SongMASS}\\
  \hline
 \textbf{\emph{Lyric}} & & \\
 \hline
 Listenability & 1.67 $\pm$ 0.62 & 2.00 $\pm$ 0.65 \\
 Grammaticality & 3.00 $\pm$ 0.76 & 3.27 $\pm$ 0.59 \\
 Meaning & 2.20 $\pm$ 0.68 & 3.20 $\pm$ 0.68 \\
 Quality & 2.27 $\pm$ 0.46 & 3.00 $\pm$ 0.38 \\
 \hline
 \textbf{\emph{Melody}} & & \\
 \hline
 Emotion & 2.40 $\pm$ 1.06 & 3.53 $\pm$ 0.64 \\
 Rhythm & 2.33 $\pm$ 1.18  & 2.87 $\pm$ 0.74  \\
 Quality & 2.33 $\pm$ 1.05  & 2.93 $\pm$ 0.70 \\
\bottomrule
\end{tabular}
\caption{\label{human_evaluation} Subjective evaluation results. Average scores and standard deviations are shown for each measure.}
\end{table}

\subsection{Results}
The main results of the objective evaluation of lyric-to-melody and melody-to-lyric generations are shown in Table~\ref{main_result}. The baseline model uses the same model structure with SongMASS, but does not leverage unsupervised melody and lyric data for pre-training and does not leverage attention-based alignment constraints. It can be seen that SongMASS greatly outperforms the baseline model in all objective metrics. The subjective evaluations are shown in Table \ref{human_evaluation}, from which we can see that the lyrics and melodies generated by SongMASS obtain better average scores in all subjective metrics. These results demonstrate the effectiveness of SongMASS in generating high-quality lyric and melody\footnote{Melody and lyric samples are available at: \url{https://musicgeneration.github.io/SongMASS/}}. We further conduct ablation study to verify the effectiveness of pre-training and alignment constraint in SongMASS. As shown in Table~\ref{main_result}, removing each component results in worse performance than SongMASS\footnote{Removing alignment constraint causes slightly better performance in PPL, which indicates that attention constraint may harm the fitting capability of the model, but still result in better generation accuracy in terms of PD, DD and MD. We also demonstrate in Table~\ref{table_align} that alignment constraint indeed improves the alignment accuracy of the generated results.}, demonstrating the contribution of pre-training and alignment constraint.

\subsection{Method Analysis}
\label{sec_method_analysis}
\paragraph{Pre-training Method}
We further investigate the effectiveness of each design in pre-training method, including using separate encoder-decoder for lyric-to-lyric and melody-to-melody pre-training and using supervised pre-training to learn a shared latent space between lyric and melody. From Table~\ref{main_result}, removing separate encoder-decoder (i.e., using shared encoder-decoder) and removing supervised loss both result in worse performance than SongMASS, which demonstrates the effectiveness of the two designs.

\begin{table}[h]
\centering
\begin{tabular}{lcc}
    \toprule
    & L2M Acc $\uparrow$ & M2L Acc $\uparrow$ \\
    \hline
    \textbf{SongMASS} & 62.6 & 45.4 \\
    - TC & 62.1 & 44.8 \\
    - SC & 56.2 & 44.0 \\ 
    - TC - SC & 55.3 & 43.8 \\
    - TC - SC - PT & 48.3 & 37.1 \\
    - DP & 15.7 & 11.3 \\
    \bottomrule
\end{tabular}
\caption{Analyses of the designs in alignment constraints.}
\label{table_align}
\end{table}

\paragraph{Alignment Strategy} We study the effectiveness of the sentence-level and token-level alignment constraints (denoted as SC and TC respectively) on the alignment accuracy (denoted in Section~\ref{sec_metric}) between melodies and lyrics. The results are shown in Table~\ref{table_align}. It can be seen that both token-level and sentence-level (especially sentence-level) constraints can improve alignment accuracy. 
It is interesting that pre-training (PT) also benefits alignment, which is probably because the patterns of lyrics and melodies are better captured with pre-training. Finally, we investigate the alignment accuracy without dynamic programming (DP) algorithm. In this case, we implement a naive alignment algorithm on attention weight matrix, which greedily decides to add another token to the current one-to-many or many-to-one alignment or to start a new alignment pair at each time step. When the sequence reaches the last token, we align all the remaining tokens of the other sequence to that token to ensure all tokens are aligned. We find that the alignment accuracy is drastically decreased without DP in Table~\ref{table_align}, showing the importance of DP for accurate alignments.

\paragraph{Alignment Visualization} To better highlight the advantages of our alignment strategy, we further visualize some cases from the lyric-to-melody tasks, as shown in Figure~\ref{song_visualization}. Figure~\ref{Song_level_vis} shows the attention weights of the whole song with and without sentence-level alignment constraints. It can be seen that the attention weights without sentence-level constrains are dispersed in all positions of the whole long sequence, and the target token cannot attend to the correct source sentences. When using sentence-level constraints, there are monotonous alignments between source and target sequence, which demonstrates the effectiveness of sentence-level alignments. Figure~\ref{Sentence_level_vis} shows the differences of whether using token-level constraints or not. We find that the attention distributions without the token-level constraints are chaotic. When applying token-level attention constraints, there are obvious diagonal trend in the attention weights, which further enable the dynamic programming algorithm to find a better alignment path as marked in red rectangles. These results demonstrate the effectiveness of token-level alignment constraints.

\section{Conclusion}
In this paper, we have proposed SongMASS, an automatic song writing system for both lyric-to-melody and melody-to-lyric generation, which leverages masked sequence to sequence pre-training and attention-based alignment constraint. We introduce some specific designs based on MASS for lyric-to-lyric and melody-to-melody pre-training, including song-level unsupervised pre-training and supervised pre-training loss to learn a shared latent space between lyric and melody. Furthermore, we introduce the sentence-level and token-level alignment constraints, and a dynamic programming algorithm to obtain accurate alignments between lyric and melody. Experimental results show that our proposed SongMASS greatly improves the quality of lyric-to-melody and melody-to-lyric generation compared with the baseline. For future work, we will investigate other sequence to sequence pre-training methods and more advanced alignment algorithms for lyric-to-melody and melody-to-lyric generation.

\section*{Acknowledgments}
This research was supported by the National Key Research And Development Program of China (No.2019YFB1405802).

\bibliography{main}

\section*{Appendix}

\subsection*{Visualization Examples}
\begin{figure}[h]
    \centering
    \subfigure[]{
        \includegraphics[width=0.48\textwidth]{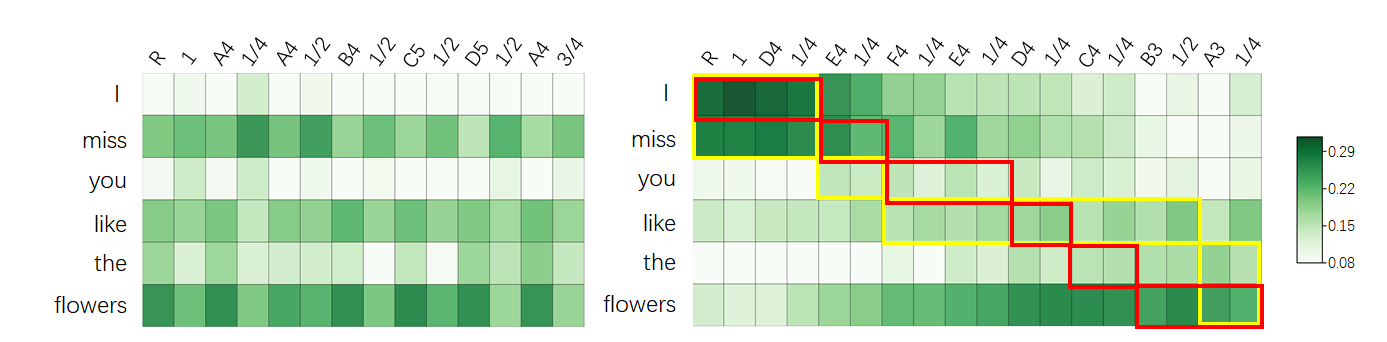}
    }
    \subfigure[]{
        \includegraphics[width=0.48\textwidth]{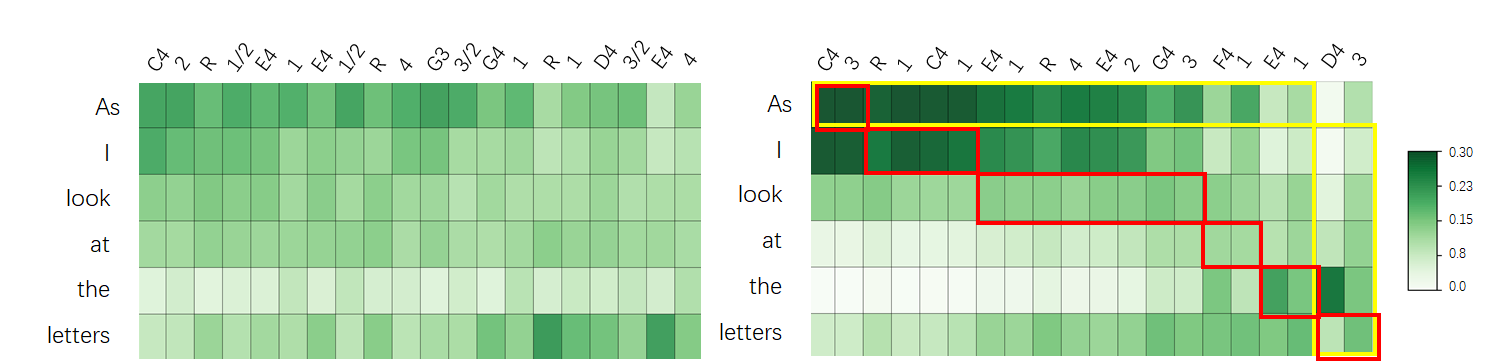}
    }
    \subfigure[]{
        \includegraphics[width=0.48\textwidth]{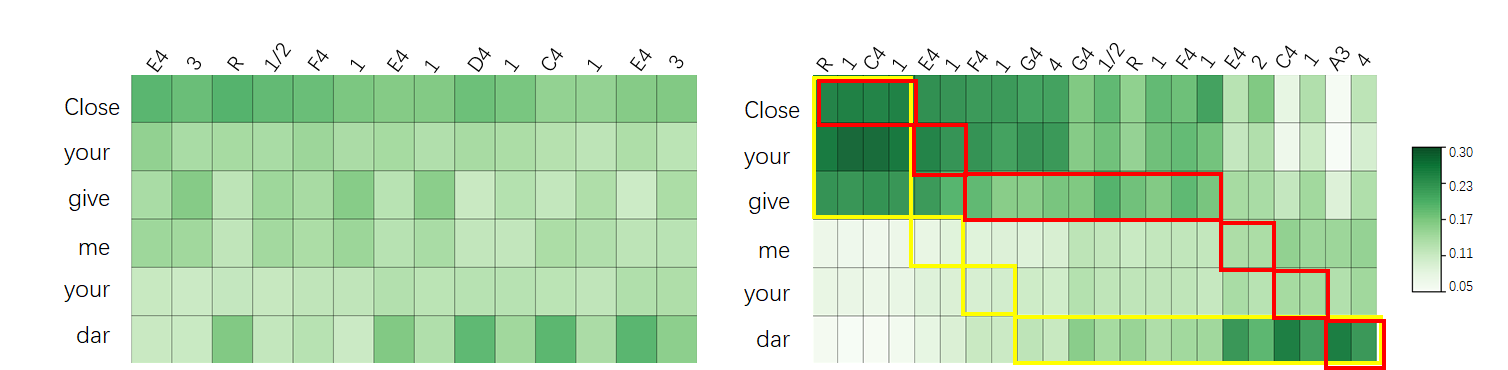}
    }
    \caption{Attention visualization. The meaning of red and yellow blocks are same as Figure~\ref{attn_example}.}
    \label{attn_example}
\end{figure}

To demonstrate the advantages of our methods in alignment, we further randomly choose some cases for attention visualization. The results are shown Figure~\ref{attn_example}. From Figure~\ref{attn_example}, we observe obvious monotonous alignments in each case, and the dynamic programming algorithm achieves more precise alignments than greedy alignments. 

\end{document}